\documentclass[runningheads]{llncs}
\usepackage[T1]{fontenc}
\usepackage{newtxtext}
\usepackage[varvw]{newtxmath}
\usepackage{graphicx}
\usepackage{booktabs}
\usepackage{xcolor}
\usepackage{hyperref}

\usepackage{tikz}
\usetikzlibrary{arrows.meta}
\begin{document}

\title{Tailoring EU AI Act Requirements from a System's Documentation: Evaluating TERE4AI on Two Open Source Systems}
\titlerunning{Tailoring EU AI Act Requirements from a System's Documentation}

\author{Jos\'e Siqueira de Cerqueira\inst{1}\orcidID{0000-0002-8143-1042} \and
Kai-Kristian Kemell\inst{1}\orcidID{0000-0002-0225-4560} \and
Jo\~ao Rossi\inst{2}\orcidID{0009-0001-3675-5005} \and
Edna Dias Canedo\inst{3}\orcidID{0000-0002-2159-339X} \and
Rebekah Rousi\inst{4}\orcidID{0000-0001-5771-3528} \and
Nannan Xi\inst{1}\orcidID{0000-0002-9424-8116} \and
Juho Hamari\inst{1}\orcidID{0000-0002-6573-588X} \and
Pekka Abrahamsson\inst{1}\orcidID{0000-0002-4360-2226}}
\authorrunning{J. Siqueira de Cerqueira et al.}
\institute{Faculty of Information Technology and Communication Sciences,
Tampere University, Tampere, Finland
\email{jose.siqueiradecerqueira@tuni.fi} \and
Department of Computer Science, University of Bras\'ilia (UnB), Bras\'ilia, Brazil \and
Computer Science Center (CIn), Federal University of Pernambuco (UFPE), Recife, PE, Brazil \and
University of Vaasa, Vaasa, Finland}

\maketitle
\setcounter{footnote}{0}

\begin{abstract}
\textbf{[Context and motivation]} Supporting regulatory compliance is a challenge for requirements engineering, yet translating regulations into software requirements remains largely manual. The EU AI Act sets requirements for high-risk AI systems that demand legal and technical knowledge.

\textbf{[Question/problem]} We study how well a tool tailors the requirements of the Act to a system from its documentation.

\textbf{[Principal ideas/results]} We evaluate TERE4AI, an open source tool that holds the Act as a knowledge graph. From a system's documentation, it proposes facts, derives the risk level and generates tailored requirements. Each cites the Act and inherits the aligned requirements of the High-Level Expert Group on AI (HLEG). For two open source systems, five AI ethics experts rated 37 sampled tailored requirements. The majority rated nearly all as correct, complete, clear and specific to the system, and agreed with their HLEG requirements. A judge model agreed with the experts' majority on correctness and completeness for all 37, and less often on clarity and specificity.

\textbf{[Contribution]} The results indicate that TERE4AI can generate requirements consistent with the Act's cited wording and specific to a system. We hope it provides developers without legal expertise with a means of translating regulations into requirements for their systems.
\end{abstract}

\section{Introduction}
\label{sec:intro}

Regulations increasingly govern how software systems are developed, deployed and used. Supporting the regulatory compliance of software is regarded as a central challenge for requirements engineering (RE)~\cite{nair2026reg2req}. Regulations, however, are written for legal audiences and not for requirements engineers, and translating them into software requirements remains largely manual and error-prone in practice~\cite{nair2026reg2req}. Several challenges complicate the translation of regulations into software requirements and the implementation of compliance, such as ambiguous legal language, cross-references, frequent amendments, and the cost and effort involved~\cite{kosenkov2025mapping,otto2007addressing}. In addition, tool support is lacking, and most existing solutions appear not to have been thoroughly assessed~\cite{kosenkov2025mapping}. Any system that supports requirements engineers must also maintain traceability from the originating law to the relevant requirements~\cite{otto2007addressing}.

The requirements of the EU AI Act are numerous and complex, and often require additional interpretation and implementation before compliance is achieved~\cite{galli2026approaching}. Extracting and classifying its obligations automatically enables organisations to determine the actions required for compliance~\cite{galli2026approaching}. Compliance with the technical documentation requirements demands advanced knowledge of both legal and technical aspects, which is rare among software developers and legal professionals~\cite{sovrano2025simplifying}. Small and medium-sized enterprises may consequently face high costs in meeting these requirements, to which independent developers are equally subject~\cite{sovrano2025simplifying}. In addition to its binding provisions, the Act refers to seven non-binding ethical principles for AI developed by the High-Level Expert Group on AI (HLEG) (recital 27)~\cite{act2024eu}. The HLEG sets these out as seven requirements, referred to in this paper as the HLEG requirements~\cite{hleg2019}. They are (1) human agency and oversight, (2) technical robustness and safety, (3) privacy and data governance, (4) transparency, (5) diversity, non-discrimination and fairness, (6) societal and environmental well-being, and (7) accountability~\cite{hleg2019}.

Considerable progress has been made in RE on the automated processing of legal text, although important limitations remain~\cite{hassani2024rethinking}. Recent approaches based on large language models (LLMs) extract the obligations of the Act~\cite{galli2026approaching,raulino2025lost}, or derive from regulatory text requirements that are independent of any particular system~\cite{abualhaija2025xtrareg,nair2026reg2req}. Other approaches assess the documentation of one system against a checklist of questions derived from the Act and return compliance scores~\cite{sovrano2025simplifying}. One such study reports that the LLM frequently generated generic requirements~\cite{abualhaija2025xtrareg}. It also notes that implementable requirements need the exact legal details to be defined for each application context~\cite{abualhaija2025xtrareg}.

This paper addresses the gap between regulations and requirements at the level of a specific system, to which the requirements of the Act are tailored from its documentation. The Act provides the basis for this tailoring, since under Article 8(1) high-risk AI systems must comply with the requirements taking into account their intended purpose~\cite{act2024eu}. Article 3(12) defines that purpose by reference to the information supplied by the provider in the instructions for use, the promotional material and the technical documentation~\cite{act2024eu}. Requirements that cite the provisions on which they rest remain traceable to them, which matters because legal traceability involves priorities and exceptions that govern special cases~\cite{otto2007addressing}. Generated requirements nevertheless require human review, since LLMs are best used as assistants to human experts~\cite{abualhaija2025xtrareg}. Under Article 16(a), the provider also remains responsible for ensuring that a high-risk system complies with the requirements~\cite{act2024eu}. The output of the tool is therefore assessed by experts in this study. This follows the guidelines for empirical studies involving LLMs for cases in which no benchmark adequately captures what is measured~\cite{baltes2025guidelines}.

Accordingly, we evaluate TERE4AI (Trustworthy Ethical Requirements Engineering for AI)~\cite{cerqueira2026tere4ai}. It represents the AI Act~\cite{act2024eu} as a knowledge graph that links each legal unit to the location of its wording and to its extracted norms. From the documentation of a system, the tool proposes facts, derives the risk level and generates tailored requirements, which cite norms and inherit the HLEG requirements aligned to them. We applied it to two public open source repositories, and five AI ethics experts assessed the same sample of its output. The study addresses three research questions.

\begin{description}
\item[RQ1] How well does the tool turn the Act into tailored requirements using a system's documentation?
\item[RQ2] To what extent do the tailored requirements support the HLEG requirements inherited from their norms?
\item[RQ3] How does the judge model's scoring compare with the experts' majority on the same tailored requirements?
\end{description}

This paper makes three contributions. First, it presents the current version of TERE4AI, an open source tool that tailors the requirements of the Act to a specific system and keeps them traceable to the Act. Second, it reports an evaluation of the tool on two public open source systems, with five AI ethics experts and a judge model. Third, it provides the 250 tailored requirements generated for the two systems, with the HLEG requirements aligned to them and the experts' ratings of a sample. We highlight that this study makes no claim of legal compliance, and that the tool neither certifies a system nor replaces a legal expert.

\section{Background and Related Work}
\label{sec:background}
\label{sec:related}

The AI Act~\cite{act2024eu} prohibits some practices (Article 5), sets requirements for high-risk systems, and adds transparency obligations for some others (Article 50). A system is high risk if it is, or is a safety component of, a product under the Annex I legislation that requires a third-party conformity assessment (Article 6(1)). It is also high risk if it falls in one of the areas of Annex III, such as employment in point 4 (Article 6(2)). Chapter III Section 2 holds the requirements of the high-risk system itself (Articles 9 to 15), and the provider must ensure that the system meets them (Article 16(a)).

The technical documentation of a high-risk system must demonstrate that the system complies with the requirements of Chapter III Section 2~\cite{omnibus2026}. The provider must also ensure that the system undergoes a conformity assessment before it is placed on the market (Article 16(f)). Kosenkov et al.\ define regulatory compliance as the state in which the functional and non-functional properties of a software-intensive product or service verifiably conform to criteria laid out in regulations~\cite{kosenkov2025mapping}. Of the 255 primary studies in their mapping study that report a challenge, 96 name the abstractness of regulations~\cite{kosenkov2025mapping}. Abstractness can prevent the immediate application of regulations as software requirements or their translation into software requirements~\cite{kosenkov2025mapping}. Otto and Ant\'on require of tools that support this translation that users can ``view the original regulatory text at any time''~\cite{otto2007addressing}. They also require that any machine-readable format stays traceable to the original wording~\cite{otto2007addressing}.

Related work includes several tools that turn a regulation into structured content. Galli et al.\ built a pipeline that extracts the AI Act's obligations into a searchable knowledge graph~\cite{galli2026approaching}. Dal Pont et al.\ extract obligations from the GDPR, the Digital Services Act and the AI Act with a similar pipeline~\cite{raulino2025lost}. XTRAREG generates privacy requirements from the GDPR, each with its legal reference~\cite{abualhaija2025xtrareg}, whereas Reg2Req derives system-agnostic requirements from the clauses of a regulation, traceable to their legal sources~\cite{nair2026reg2req}. Other approaches start from one system. Sovrano et al.\ use LLM pipelines to help developers assess and write the technical documentation of a high-risk system under the Act~\cite{sovrano2025simplifying}. Z-Inspection is a process in which experts assess one AI system and link its ethical issues to the HLEG requirements~\cite{zicari2021zinspection}. TERE4AI differs from all of them, since it derives the risk level of one system from facts in its documentation. It then generates tailored requirements for that system, each citing norms of the Act (Section~\ref{sec:tool}).

\section{Current Version of TERE4AI}
\label{sec:tool}

TERE4AI is an open source Model Context Protocol (MCP) server that holds the EU AI Act as a knowledge graph. MCP is an open protocol through which applications built on LLMs connect to external data sources and tools. A coding agent can thus call TERE4AI while it builds a system. The same functions are also reachable over HTTP, which TERE4AI-web, the web client used in this study, calls. TERE4AI-web receives the address of a public repository and returns the proposed facts, the risk level and the tailored requirements with their aligned HLEG requirements. A person can thus obtain them for any public repository without a coding agent. An earlier version generated requirements from a description of the system typed by the user~\cite{cerqueira2026tere4ai}. The current version instead reads the system's documentation and shows each proposed fact to a person, who can change it before a deterministic script decides the risk level.

\begin{figure}[t]%
  \centering
  \includegraphics[width=0.36\textwidth]{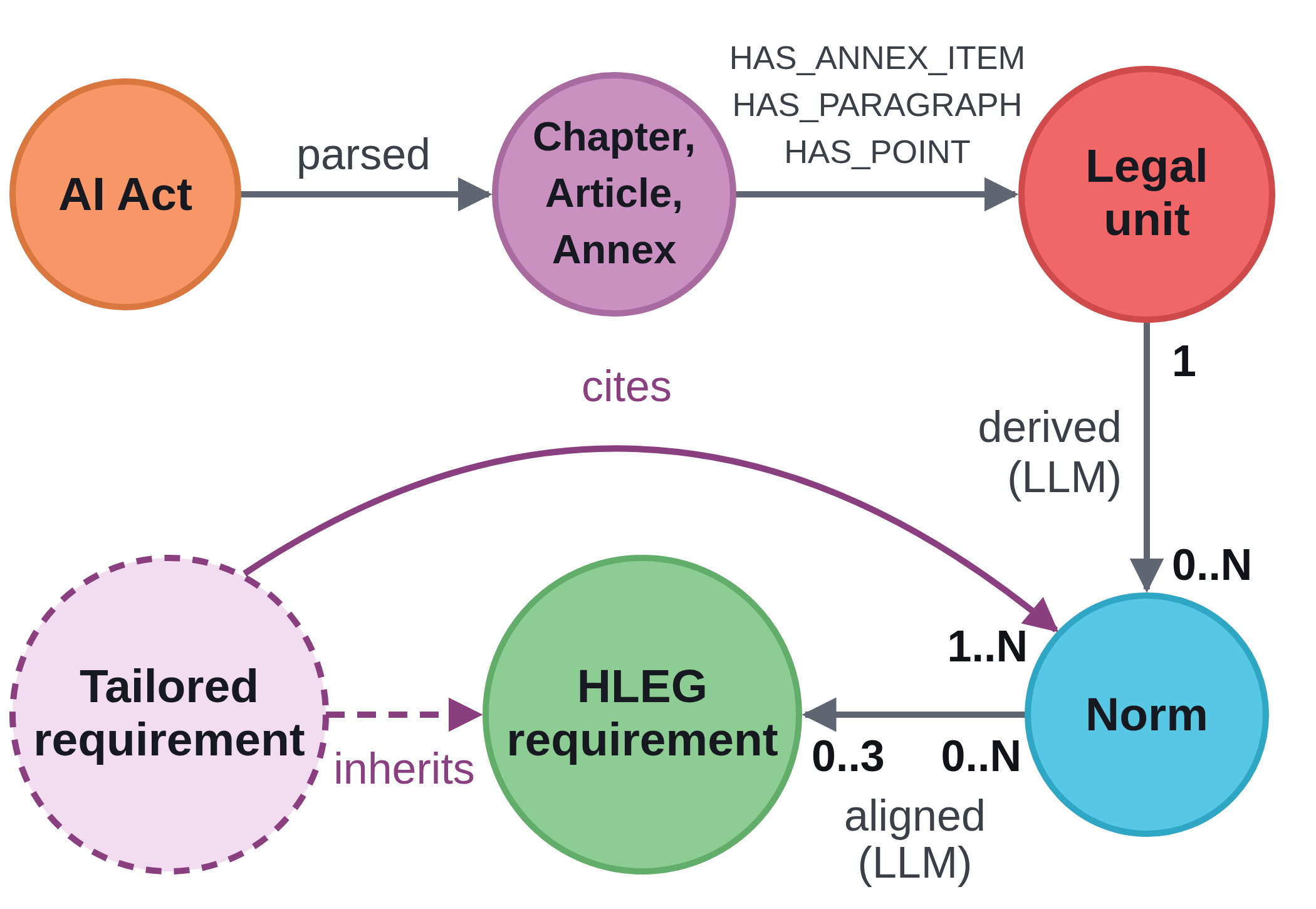}%
  \hspace{0.02\textwidth}%
  \includegraphics[width=0.36\textwidth]{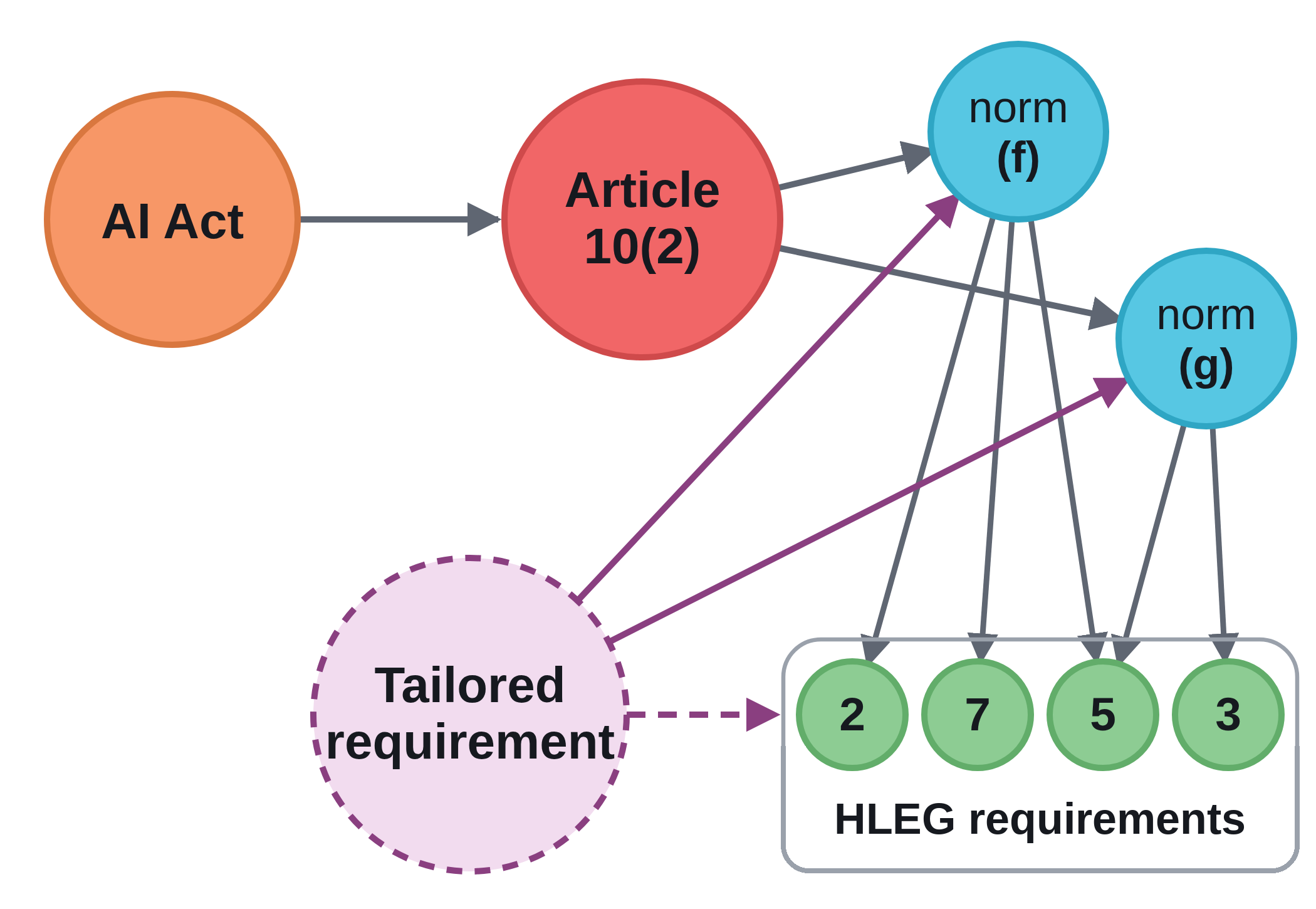}%
  \caption{The knowledge graph of TERE4AI. Left: legal units, norms and their alignments to HLEG requirements are stored at build time, whereas a tailored requirement (dashed) is generated at use time. Right: a tailored requirement that cites two norms of Article~10(2) and inherits their HLEG requirements.}%
  \label{fig:graph}%
\end{figure}

Figure~\ref{fig:graph} (left) shows the knowledge graph, which is built once, before any system is seen. The Act, as amended by Regulation (EU) 2026/1744~\cite{omnibus2026}, is parsed by a script, without the use of an LLM, into legal units. These are Articles, paragraphs, points, annex items and recitals, each keeping the location of its wording. A legal unit is often too large to be translated into a requirement, since one paragraph can state several duties. Earlier approaches therefore extract obligations at the level of the sentence and represent each by elements such as its deontic modality, its addressee and its pre-conditions~\cite{galli2026approaching,raulino2025lost}. TERE4AI follows them, and a generator model extracts norms from a core of 30 Articles and Annexes. We use the term norm for one obligation, prohibition, permission, exemption or right extracted in this way. A judge model accepts, rejects or holds each norm for review, and only accepted norms are used. It belongs to another model family than the generator, since a model's rating can be inflated when the model is similar to the evaluator~\cite{panickssery2024llm}. The generator then aligns each accepted norm to at most three of the seven HLEG requirements, and the judge accepts or rejects each alignment. One complete result of these steps is called a build. The build used in this study, generated by \texttt{gpt-6-astra} and judged by \texttt{claude-opus-5-5}, holds 537 accepted alignments, which are not validated by experts. It is available in the replication package.

At use time, the web client selects the documentation files of a repository by the reading rule of Section~\ref{sec:reading}. One generator call then proposes values for the 44 facts that the script reads, such as the domain, the autonomy and flags that mirror Article 5. A fact is kept only if its quote is found in the text read, and the web client links each quote to a file and a line. Before the risk level is decided, a person can keep, change, reject or add a fact. The script then derives the risk level from the confirmed facts, and returns the rule that decided it and the facts left unknown. For each applicable Article, one generator call receives the Article's accepted norms together with the system's context. It generates tailored requirements, each with a priority and the norms it cites. The judge model then checks each Article's tailored requirements against the norms. This check is the gate of the tool, and a rejected generation, that is, the tailored requirements of one Article, is marked for human review. Finally, each tailored requirement inherits, without the use of an LLM, the HLEG requirements aligned to the norms it cites.
In the example of Figure~\ref{fig:graph} (right), Article~10(2) is one legal unit with ten norms, two of which concern examining the data for possible biases and measures against them. A tailored requirement that cites these two norms inherits the four HLEG requirements to which they are aligned.

\section{Study Design}
\label{sec:design}

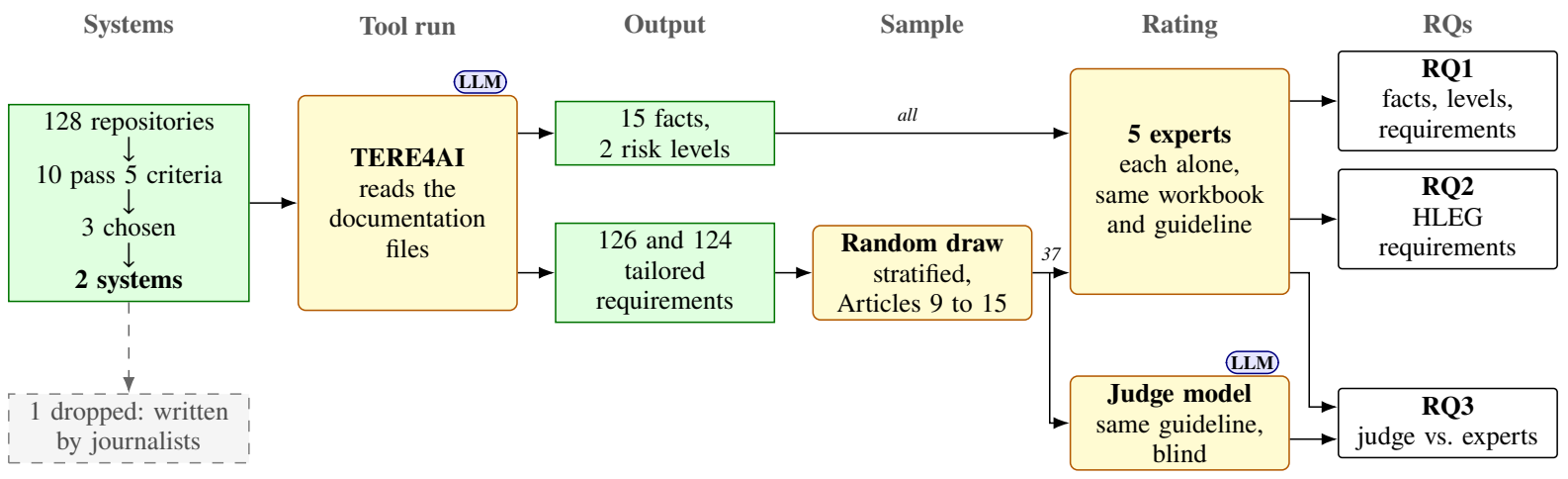
\begin{figure}[t]
  \centering
  \resizebox{0.85\textwidth}{!}{%
\begin{tikzpicture}[font=\scriptsize, >={Latex[length=1.5mm]},
  set/.style={draw=green!45!black, fill=green!12, align=center, inner sep=2pt},
  stp/.style={draw=orange!70!black, fill=yellow!22, align=center, inner sep=2pt, rounded corners=2pt},
  rq/.style={draw=black, align=center, inner sep=2pt, rounded corners=1pt, text width=19mm},
  badge/.style={draw=blue!50!black, fill=blue!10, rounded corners=3pt, inner sep=1.2pt, font=\tiny\bfseries},
  head/.style={font=\scriptsize\bfseries, text=black!65}]
\node[set, text width=21mm] (pool) at (1.15,0.1) {128 repositories\\[-1pt]$\downarrow$\\[-1pt]10 pass 5 criteria\\[-1pt]$\downarrow$\\[-1pt]3 chosen\\[-1pt]$\downarrow$\\[-1pt]\textbf{2 systems}};
\node[draw=black!50, dashed, fill=black!4, align=center, inner sep=2pt, text width=21mm, text=black!70] (drop) at (1.15,-2.0) {1 dropped: written\\by journalists};
\node[stp, text width=19mm, minimum height=20mm] (tool) at (3.75,0.1) {\textbf{TERE4AI}\\reads the\\documentation\\files};
\node[badge] at ([xshift=-3.5mm,yshift=1.2mm]tool.north east) {LLM};
\node[set, text width=19mm] (facts) at (6.15,0.75) {15 facts,\\2 risk levels};
\node[set, text width=19mm] (reqs) at (6.15,-0.55) {126 and 124\\tailored\\requirements};
\node[stp, text width=19mm] (draw) at (8.55,-0.55) {\textbf{Random draw}\\stratified,\\Articles 9 to 15};
\node[stp, text width=19mm, minimum height=21mm] (exp) at (10.95,0.3) {\textbf{5 experts}\\each alone,\\same workbook\\and guideline};
\node[stp, text width=19mm] (judge) at (10.95,-1.95) {\textbf{Judge model}\\same guideline,\\blind};
\node[badge] at ([xshift=-3.5mm,yshift=1.2mm]judge.north east) {LLM};
\node[rq] (rq1) at (13.45,1.05) {\textbf{RQ1}\\facts, levels,\\requirements};
\node[rq] (rq2) at (13.45,-0.05) {\textbf{RQ2}\\HLEG\\requirements};
\node[rq] (rq3) at (13.45,-1.95) {\textbf{RQ3}\\judge vs.\ experts};
\draw[->, dashed, black!60] (pool) -- (drop);
\draw[->] (pool.east |- tool) -- (tool);
\draw[->] (tool.east |- facts) -- (facts);
\draw[->] (tool.east |- reqs) -- (reqs);
\draw[->] (facts.east) -- node[above, font=\tiny\itshape, pos=0.45] {all} (facts.east -| exp.west);
\draw[->] (reqs) -- (draw);
\draw[->] (draw.east) -- node[above, font=\tiny\itshape, pos=0.5] {37} (draw.east -| exp.west);
\draw[->] ([xshift=1.6mm]draw.east) |- (judge.west);
\draw[->] (exp.east |- rq1) -- (rq1);
\draw[->] (exp.east |- rq2) -- (rq2);
\draw[->] ([yshift=-1.5mm]judge.east) -- ([yshift=-1.5mm]rq3.west);
\draw[->] ([yshift=-8.5mm]exp.east) -- ++(1.8mm,0) |- ([yshift=1.5mm]rq3.west);
\foreach \n/\t in {pool/Systems, tool/Tool run, facts/Output, draw/Sample, exp/Rating, rq1/RQs}
  \node[head] at (\n |- 0,1.75) {\t};
\end{tikzpicture}}
  \caption{Study design. Green boxes are sets with their size, yellow boxes are steps, and LLM marks a step performed by an LLM.}
  \label{fig:design}
\end{figure}

Following an adapted Design Science Research (DSR) approach~\cite{hevner2004design}, TERE4AI is built iteratively in four phases, namely exploration, prototyping, evaluation and communication of results. In the exploration phase, earlier studies identified that enterprises require adequate tools to comply with the Act~\cite{sillberg2025falls}. They also suggested the need to combine technical interventions, such as knowledge graphs, with governance mechanisms, such as human-in-the-loop, to enhance the trustworthiness of LLMs~\cite{cerqueira2025mapping}. The prototyping phase resulted first in an LLM-based multi-agent system~\cite{cerqueira2025trust} and then in an earlier version of TERE4AI, which was demonstrated but not evaluated with practitioners~\cite{cerqueira2026tere4ai}. In this study, we focus on the evaluation phase for the current version of TERE4AI, described in Section~\ref{sec:tool}. It is a case study-based evaluation that follows the guidelines for case study research~\cite{runeson2009guidelines}. The tool is applied to two open source systems selected on purpose, and five AI ethics experts assess its output. Figure~\ref{fig:design} shows the design, from the selection of the systems to the research questions. The study is reported against the guidelines for empirical studies involving LLMs~\cite{baltes2025guidelines} and, for the ratings, the inter-rater reliability and agreement supplement of the ACM SIGSOFT Empirical Standards~\cite{ralph2026irr}. The generator is \texttt{gpt-6-astra} and the judge model \texttt{claude-opus-5-5}, both at effort high.

The systems were selected intentionally and not sampled~\cite{runeson2009guidelines}. Since systems placed on the market as high risk are often proprietary, we searched GitHub, where public repositories hold the documentation of a system beside its code. In an exploratory search, we assembled an initial set of 128 public repositories of AI systems in the areas of the Act. The repositories and each step of the selection are available in the replication package. Five criteria were applied in order. A repository had to be an application and not a library, state the facts that decide the risk level in its own files, and name an organisation that ran the system. It also had to hold its documentation, and the selection had to vary across areas. Ten repositories passed, and we chose three in three Annex III areas. The third, a welfare-fraud risk scoring system, was dropped before the runs because its documentation was written by journalists and not by its provider. The first remaining system is a resume-ranking tool for hiring\footnote{\url{https://github.com/interviewstreet/hiring-agent/tree/e0ece80}, the commit that was run.}, which the tool classified under employment (Annex III point 4). The second is a medical image segmentation tool that also predicts body statistics from scans\footnote{\url{https://github.com/wasserth/TotalSegmentator/tree/c59be95}, the commit that was run.}, classified under biometric categorisation (point 1) and chosen because its deciding fact is contestable. We treat both systems as if they were placed on the market. The study does not decide whether they are covered in law, and the tool's classification is itself one of the outputs the experts assess. Each repository was run end to end on the build \texttt{2026-10-16-gpt-6-astra-high-opus-5-5} on 4 November 2026, with the facts confirmed as proposed.

\label{sec:reading}
The tool reads only the documentation of a repository and never its code: its READMEs, model cards, dataset cards, system documents such as an impact assessment, and text files under a documentation folder.

The two runs yielded 126 and 124 tailored requirements over 24 Articles (Table~\ref{tab:read}), more than the experts could rate. The rated set is therefore restricted to Articles 9 to 15, the requirements for the high-risk system itself, which apply irrespective of its area. It is a stratified random sample of 37 tailored requirements, 18 for the hiring tool and 19 for the segmentation tool. The strata are the 14 groups of Article by system, and three tailored requirements were drawn from each, or all where an Article yielded fewer. A random draw is what the agreement coefficient assumes of its units~\cite{hayes2007answering}. All experts rated the same rows, so that a majority and an agreement coefficient exist. They also rated every proposed fact of the two systems, 15 in all, and the two risk levels.

Each expert received one workbook with identical rows and a written guideline, both adapted from the Stage 2 annotation guideline of Reg2Req~\cite{nair2026reg2req}. For each tailored requirement, the expert saw its text, the norms it cites and the wording of the Act for each norm. The expert also saw a summary of the system's facts and the HLEG requirements inherited from the norms. The expert rated correctness, completeness and clarity from 1 to 5, with every level anchored in words. Unlike in Reg2Req, completeness was rated for each tailored requirement against the norms it cites, not for each clause. The first of two questions that are this study's own, each answered yes or no, is whether the tailored requirement is specific to this system, which operationalises tailoring. The second, for RQ2, is whether it supports every HLEG requirement shown, in whole or in part. The guideline provides what each HLEG requirement includes and holds two worked examples drawn from rows outside the sample. A comment was required for a correctness or completeness score of 1 or 2 and for each negative answer. On two further sheets, the expert indicated whether each proposed fact is right for the repository and gave an own reading of the risk level.

Six AI ethics experts, three at Brazilian and three at Finnish universities, were recruited by email through the authors' network. Each holds or is pursuing a doctorate and has published at least one paper on AI ethics. Five experts rated both systems, whereas the sixth returned the workbook in part, with ratings for 17 of the 18 tailored requirements of the hiring tool only. After the workbooks had been returned, we set the rule that the measures rest on the five experts who rated both systems. The answers of the sixth expert are therefore reported apart from the measures. Before the experts received the workbook, three of the authors piloted ten rows, and the guideline was revised where the pilot showed a misreading. Four of the experts were introduced to the workbook in a video meeting, whereas two received the written material only. Each expert rated the workbook alone, and scores were not reconciled afterwards, since agreement among the experts is itself a result.

A score counts as high when it is 4 or 5, or yes for the two questions, and this is the threshold. For RQ1, the measure for the facts is the share of values that the majority found right, and for the risk level the experts' readings are set against the tool's. For the tailored requirements, the measure for each criterion is the number of the 37 for which the majority, at least three of the five experts, is at or above the threshold. Inter-rater agreement among the five experts is Krippendorff's alpha, ordinal for the scores and nominal for the yes or no answers, with a 95 percent bootstrap interval~\cite{hayes2007answering}. Alpha is reported for each criterion~\cite{ralph2026irr}, together with the share of expert pairs on the same side of the threshold. For RQ2, the measure is the number of the 37 whose majority answered yes, with alpha as above. For RQ3, following the guidelines for subjective judgments~\cite{baltes2025guidelines}, the judge model rated the same 37 tailored requirements in one run under the same guideline. It saw neither the experts' answers nor the gate's verdict. It is compared with the experts' majority and with each expert, and is reported apart from the experts. For each criterion, the comparison gives how often the judge model and the majority are on the same side of the threshold. A false accept is a tailored requirement that the judge model placed at or above the threshold and the majority below it, and a false reject is the reverse. Alpha is also given between the judge model and the experts' reference, which is their median for the scores and their majority for the specificity question. A tailored requirement without a majority is left out of the comparison.

\section{Results}
\label{sec:results}

\subsection{RQ1: How well does the tool turn the Act into tailored requirements using a system's documentation?}

The tool read 3 and 2 documentation files (Table~\ref{tab:read}) and proposed 15 facts in all, each with a quote located in a file and a line. Of the 44 facts of the schema, 38 and 39 remained unknown. The majority of the experts found all 15 proposed values right, and 62 of the 72 answers given were yes (alpha $-0.03$, interval $-0.13$ to 0.07). All five experts read the hiring tool as high risk, the risk level that the tool assigned. For the segmentation tool, to which the tool also assigned high risk, three experts read high risk, one read prohibited and one could not tell. Two experts, one of whom read high risk, commented that the documentation alone does not support the rule on biometric categorisation that decided this risk level.

\begin{table}[t]
\centering
\caption{What the tool read and produced for each repository. Unknown counts those of the 44 facts left without a value, and two of the facts proposed for each system are lists outside the 44.}
\label{tab:read}
\begin{tabular}{lrrrrr}
\toprule
 & Files & Facts & & Applicable & Tailored \\
Repository & read & proposed & Unknown & Articles & requirements \\
\midrule
Hiring tool & 3 & 8 & 38 & 24 & 126 \\
Image segmentation & 2 & 7 & 39 & 24 & 124 \\
\bottomrule
\end{tabular}
\end{table}

For all 37 sampled tailored requirements, at least three of the five experts rated correctness and completeness 4 or 5 (Table~\ref{tab:req}). The majority also rated 35 of the 37 at 4 or 5 on clarity and found 36 specific to the system, whereas one had no majority. The five experts gave no score of 1, and at least 95 percent of their scores on each criterion were 4 or 5. The two tailored requirements below the threshold on clarity both concern Article 13, and their descriptions are the longest in the sample, at 297 and 353 words. Six tailored requirements were marked as not specific by at least one expert, four of them with ``generic'' as the written reason. All six are built on a norm that gives an option or applies only under a stated condition. Their only two scores of 2 concern the correctness of two of these six, where one expert held that the priority ``must'' creates ``an obligation not stated''. Another expert rated the same two 5 and commented on one that the simplification is ``correctly framed as optional''.

\begin{table}[t]
\centering
\caption{Expert scores of the 37 sampled tailored requirements. Majority: tailored requirements with the majority at or above the threshold, where one had no majority on the specificity question. Per expert: the lowest and highest count at or above the threshold among the five. Alpha is given with its 95 percent bootstrap interval. Pairs: the share of expert pairs on the same side of the threshold.}
\label{tab:req}
\small
\setlength{\tabcolsep}{4pt}
\begin{tabular}{lccccc}
\toprule
Criterion & Majority & Scores 1 to 3 / 4 / 5 & Per expert & Alpha & Pairs \\
\midrule
Correctness & 37 of 37 & 9 / 64 / 111 & 33 to 37 & $-0.13$ ($-0.19$, $-0.08$) & 0.91 \\
Completeness & 37 of 37 & 6 / 62 / 116 & 31 to 37 & $-0.12$ ($-0.17$, $-0.06$) & 0.93 \\
Clarity & 35 of 37 & 9 / 126 / 48 & 33 to 36 & 0.16 (0.04, 0.27) & 0.93 \\
Specific & 36 of 37 & 9 no / 175 yes & 33 to 37 & 0.15 ($-0.01$, 0.25) & 0.92 \\
\bottomrule
\end{tabular}
\end{table}

Krippendorff's alpha was slightly below zero for correctness and completeness and low for clarity and the specificity question (Table~\ref{tab:req}). Since almost every score was 4 or 5, the scores hardly distinguish one tailored requirement from another, in which case alpha can be low despite frequent agreement~\cite{krippendorff2013commentary}. The experts also differed in how high they rated, since one rated every tailored requirement 5 on correctness, whereas two rated most of them 4. Stable differences of this kind add disagreement within each tailored requirement, the condition under which alpha falls below zero. The share of expert pairs on the same side of the threshold was 91 to 93 percent, close to what the frequency of high scores alone produces. The ratings therefore show that each expert rated the output high, although they give no evidence that the experts rank the tailored requirements alike.

\subsection{RQ2: To what extent do the tailored requirements support the HLEG requirements inherited from their norms?}

Of the 250 generated tailored requirements, 228 show at least one HLEG requirement, and the 37 sampled ones all do, with 84 HLEG requirements shown in all (Table~\ref{tab:hleg}). For 36 of the 37, the majority of the experts answered that the tailored requirement supports every HLEG requirement shown, whereas for one it answered no. Of the 178 answers given, 167 were yes and 11 were no, and two of the five experts never answered no. Alpha was 0.18 (interval $-0.02$ to 0.35), which is low for the same reason as in RQ1. The share of expert pairs giving the same answer was 90 percent, against 88 percent produced by the frequency of yes alone.

Nine of the 11 negative answers name the HLEG requirement not supported, and they fall on four of the seven. Societal and environmental well-being was rejected on both tailored requirements for Article 13, the only two in the sample that show it. One of them has the majority no, and an expert wrote there that this HLEG requirement ``is not explicit here''. In the build, societal and environmental well-being has only 2 of the 537 accepted alignments. Human agency and oversight was rejected on both tailored requirements for Article 11, because ``creating a technical dossier alone does not operationally enable or assign human oversight''. Privacy and data governance was rejected on both tailored requirements for Article 12, and diversity, non-discrimination and fairness on one for Article 10, each by one expert. In every case, the rejected HLEG requirement is inherited from exactly one of the cited norms, one of 18 for Article 13. The rejections thus mostly recur on the same norms across the two systems, whereas the cause, the alignment or the tailored text, was not tested.

\begin{table}[t]
\centering
\caption{HLEG requirements shown per system. A tailored requirement counts under each HLEG requirement shown for it. The tool shows 524 of the 537 accepted alignments, those stored as supporting or partially operationalising an HLEG requirement. They are generated by an LLM and not validated by experts.}
\label{tab:hleg}
\setlength{\tabcolsep}{4pt}
\begin{tabular}{lrrrr}
\toprule
& \multicolumn{2}{c}{All 250} & \multicolumn{2}{c}{Sampled 37} \\
HLEG requirement & Hiring & Segm. & Hiring & Segm. \\
\midrule
Human agency and oversight & 47 & 46 & 7 & 8 \\
Technical robustness and safety & 38 & 37 & 10 & 11 \\
Privacy and data governance & 18 & 20 & 6 & 6 \\
Transparency & 30 & 31 & 8 & 8 \\
Diversity, non-discrimination and fairness & 14 & 16 & 4 & 1 \\
Societal and environmental well-being & 1 & 1 & 1 & 1 \\
Accountability & 74 & 67 & 8 & 5 \\
\bottomrule
\end{tabular}
\end{table}

\subsection{RQ3: How does the judge model's scoring compare with the experts' majority on the same tailored requirements?}

The tool's own gate accepted 11 of the 14 sampled Article generations and rejected 3 (Article 9 for both systems, Article 10 for the hiring tool). The 9 sampled tailored requirements from these 3 generations were all at or above the threshold for the majority of the experts on every criterion. The judge model was on the same side of the threshold as the majority for all 37 tailored requirements on correctness and on completeness (Table~\ref{tab:judge}). On clarity it was on the same side for 31, and on the specificity question for 30 of the 36 that have a majority. A judge that rated every tailored requirement high would have matched the majority 37, 37, 35 and 36 times. The comparison therefore shows no gain of the judge over a constant answer, since the experts placed almost nothing below the threshold. Alpha between the judge and the experts' reference had intervals that include zero for the three scores. Against each expert alone, the judge was on the same side for 81 to 100 percent of the tailored requirements.

\begin{table}[t]
\centering
\caption{The judge model against the five experts on the 37 sampled tailored requirements. Same side, false accepts and false rejects are counted against the experts' majority, which one tailored requirement lacks on the specificity question. Alpha is between the judge model and the experts' reference, with its 95 percent bootstrap interval. A count of 0 of 0 means that the majority placed no tailored requirement below the threshold.}
\label{tab:judge}
\small
\setlength{\tabcolsep}{3pt}
\begin{tabular}{lccccc}
\toprule
 & Judge & Same & False & False & \\
Criterion & 1 to 3 / 4 / 5 & side & accepts & rejects & Alpha \\
\midrule
Correctness & 0 / 7 / 30 & 37 of 37 & 0 of 0 & 0 of 37 & 0.36 ($-0.09$, 0.72) \\
Completeness & 0 / 4 / 33 & 37 of 37 & 0 of 0 & 0 of 37 & 0.32 ($-0.11$, 0.72) \\
Clarity & 4 / 31 / 2 & 31 of 37 & 2 of 2 & 4 of 35 & 0.12 ($-0.07$, 0.34) \\
Specific & 7 no / 30 yes & 30 of 36 & 0 of 0 & 6 of 36 & $-0.08$ ($-0.16$, $-0.01$) \\
\bottomrule
\end{tabular}
\end{table}

On clarity, the majority placed only the two long tailored requirements for Article 13 below the threshold, and the judge rated both 4. The judge rated four others 3, where none of the five rated below 4. On correctness and completeness, the judge never rated below 4, including the two tailored requirements to which one expert gave a correctness score of 2. On the specificity question, the judge answered no for seven tailored requirements, and all six for which at least one expert answered no are among them. The seven cite a permission or a conditional duty, and against the majority six of them count as false rejects, whereas the seventh has no majority. Human-judge agreement of this kind indicates how far the judge matches the experts on these 37, not that either is right.

\section{Discussion}
\label{sec:discussion}

Among the tools discussed in Section~\ref{sec:background}, to the best of our knowledge only TERE4AI decides which part of the Act applies to a system before generating requirements for it. The five experts rated the sampled tailored requirements high on every criterion, for two systems chosen because their documentation states the deciding facts. Since almost all scores were 4 or 5, they indicate that the experts found no serious fault, but they do not distinguish the better tailored requirements from the weaker ones. A sixth expert, whose partial ratings are reported apart from the measures (Section~\ref{sec:design}), found the tailored requirements difficult to assess in isolation. This expert proposed a multiprofessional focus group of governance, legal and engineering experts.

The ratings and comments of the experts indicate that the tailored requirements follow the wording of the cited norms and refer to the purpose and components of each system. One expert, for instance, described the monitoring diagnostics of one tailored requirement as ``specific and testable''. Negative feedback was rare and mainly concerned tailored requirements generated from norms that apply only in some cases, such as testing in real-world conditions, which a provider may choose. The tool selects the Articles that apply to a system, but it does not check whether each norm of these Articles applies to it. One expert accordingly found such a tailored requirement ``generic and applicable to other AI systems''. XTRAREG's authors report generic requirements as well, in their case requirements that are not detailed beyond the legal formulation and sometimes overlook exceptions~\cite{abualhaija2025xtrareg}. One expert further noted that the applicability of some provisions to the systems as described needs confirmation, a decision that requires legal expertise. Some experts also noted that tailored requirements with subjective wording leave open how much has to be done and how fulfilment is verified. Clarity also fell as more norms were grouped into one tailored requirement, and one expert advised to ``consider splitting into verifiable sub-requirements''.

The majority of the experts accepted the HLEG requirements shown for 36 of the 37 tailored requirements, on a question that covered every HLEG requirement shown at once. All 11 negative answers concern tailored requirements that show three or more HLEG requirements. The seven rejected HLEG requirements are all inherited from a single cited norm, whereas none of the 48 inherited from two or more norms was rejected. This pattern is descriptive, since it rests on four alignments, yet it indicates which alignments experts should validate first. For societal and environmental well-being, the rejection is consistent with the texts themselves, since Article 95(2)(b) places environmental sustainability among the matters of voluntary codes of conduct~\cite{act2024eu}. The HLEG in turn states that the full realisation of ethical and robust AI may go beyond existing legal obligations~\cite{hleg2019}. The HLEG requirements here were proposed by the tool, and the experts accepted or rejected them without seeing the evidence behind each. In Z-Inspection, in contrast, experts map the ethical issues of a system to the HLEG requirements themselves and hold a consensus discussion where their scores differ by more than 2~\cite{zicari2021zinspection}.

Krippendorff's alpha among the experts was close to zero on every criterion, although almost every score was high. Galli et al.\ report negative alpha values for the Predicate and Beneficiary elements of the obligations they extract from the Act, which they read as systematic disagreement~\cite{galli2026approaching}. Here the experts differed in how high they rated, and their opposite readings on correctness and the specificity question concern tailored requirements from norms that apply only in some cases. On correctness and completeness, the judge model agreed with the experts' majority exactly as often as if it had rated every tailored requirement high. On clarity and the specificity question, it agreed less often than that. Sovrano et al.\ met the same difficulty, as a system that always answered yes achieved very high performance and the skewed answers complicated the interpretation~\cite{sovrano2025simplifying}. Six of the judge model's seven answers of not specific fall on tailored requirements that one or two experts also marked as not specific. Both may follow the guideline's second worked example, which answers not specific for a tailored requirement from a norm that applies only in some cases. The same model is the gate of the tool, where it rejected three sampled generations. Under the experts' guideline, it placed their nine sampled tailored requirements at or above the threshold on correctness and completeness. The verdict of the gate therefore cannot be read as a statement about every tailored requirement of a rejected generation. Moreover, where the judge model and the experts agree, the agreement does not imply that the measurement captures the intended construct~\cite{baltes2025guidelines}.

\begin{figure}[t]
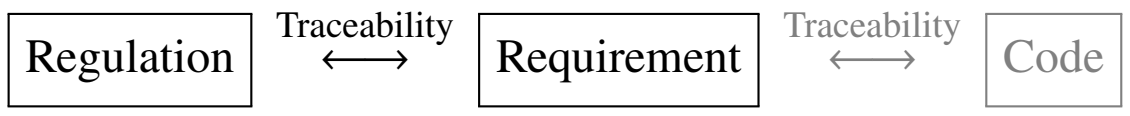

\centering
\fbox{\strut Regulation}\ \ $\stackrel{\mbox{\scriptsize Traceability}}{\longleftrightarrow}$\ \ \fbox{\strut Requirement}\ \ \textcolor{gray}{$\stackrel{\mbox{\scriptsize Traceability}}{\longleftrightarrow}$\ \ \fbox{\strut Code}}
\caption{Traceability from a regulation to code. The first traceability link was evaluated in part in this study, and the second (grey) is left to future research.}
\label{fig:trace}
\end{figure}

Figure~\ref{fig:trace} places these results on the two traceability links that connect a regulation to code. The first traceability link is the one that Otto and Ant\'on require of any tool support, namely traceability from the originating law to the requirements~\cite{otto2007addressing}. In TERE4AI, each tailored requirement cites norms from a closed set, and each norm resolves to the wording of the Act. The experts rated each tailored requirement against the wording it cites, which is one part of this traceability link. Whether the cited norms are the right ones, and whether norms are missing, was not rated. The second traceability link, from a tailored requirement to the code, remains open, since five of the 18 tailored requirements sampled for the hiring tool name Python files that no expert opened. As the sixth expert wrote, ``I can't for example know what the .py files in the requirements contain''. TERE4AI is aimed at developers and small and medium-sized enterprises without legal expertise, for whom drafting and reviewing technical documentation for compliance with the Act is challenging~\cite{sovrano2025simplifying}. They may use its output as a first set of requirements for their own system, to be reviewed for the conditions and options that the tool leaves undecided.

\section{Threats to Validity}
\label{sec:threats}

We report the threats to validity according to Wohlin et al.~\cite{wohlin2012experimentation}. On construct validity, the experts are not lawyers, so a high correctness score indicates agreement with the cited wording, not that the Act legally applies as read. Moreover, a fact not proposed, a norm not cited and an HLEG requirement not shown cannot lower a score. The specificity question combines whether a tailored requirement names the system with whether its provision applies, and its worked example may have guided experts and judge alike. On internal validity, the experts saw the tool's output and the HLEG requirements shown before answering, which may have raised their scores, most of all for RQ2. The authors chose the systems, wrote the guideline and recruited the experts through their own network.

On external validity, the study covers two open source systems, and only tailored requirements for Articles 9 to 15 were rated. One generator and one judge model, both commercial, were used on one build. The findings extend only analytically, to systems whose documentation states the deciding facts~\cite{runeson2009guidelines}. On conclusion validity, almost every score was 4 or 5, so the ratings neither rank the tailored requirements nor show how often the judge model misses a weak one. The majority as reference also hides the minority readings, and scores were not reconciled by discussion. RQ3 rests on at most 37 comparisons of the judge model with the experts, fewer than the 100 that related fields commonly use~\cite{baltes2025guidelines}. The models' nondeterminism is not sampled, although the guidelines recommend repeated runs~\cite{baltes2025guidelines}. No baseline was run, although the guidelines recommend benchmarking a tool against suitable baselines and using an open model as a baseline whenever it is technically feasible~\cite{baltes2025guidelines}. The rule that leaves the sixth expert out of the measures was set after the workbooks had been returned. This expert rated clarity below 4 for 9 of 17 tailored requirements, against at most one of the same 17 for each of the five. With that expert's answers added, no majority on a tailored requirement changes on any criterion, and alpha moves by at most 0.03.

\section{Conclusion and Future Work}
\label{sec:conclusion}

TERE4AI is an open source tool that provides requirements grounded in the EU AI Act and aligned with the HLEG requirements. This study evaluated its use through the web client TERE4AI-web, which tailors these requirements to a system from the documentation in its repository. Five AI ethics experts and a judge model assessed the tailored requirements generated for two open source systems. Our results indicate that TERE4AI can generate requirements that are consistent with the cited wording of the Act and specific to a system, as described in its documentation. The majority of the experts rated nearly all tailored requirements high on every criterion and agreed with their HLEG requirements. Nevertheless, the experts indicated that some tailored requirements combine several requirements into one, and that the applicability of some norms requires legal expertise. The judge model agreed with the experts on most tailored requirements, although no more often than if it had rated every one high. These results suggest that TERE4AI has the potential to support developers without legal expertise in translating regulations into traceable requirements. As we evaluated two open source systems with proprietary LLMs, we plan a broader study with systems in other domains, more participants, open-weight models and a baseline. We also plan to study the traceability between tailored requirements and code when coding agents use the tool in different contexts. Furthermore, the norms and alignments held in the knowledge graph should be validated by experts. The tool should also be evaluated in a focus group with legal experts, AI ethicists and developers. Finally, adapting the tool to other regulations and to industrial standards is worth exploring.

\subsubsection*{Data Availability Statement.} The replication package is available at \url{https://doi.org/10.5281/zenodo.24341061}. It holds the source of TERE4AI and of TERE4AI-web, the build and the prompts, the study materials, the ratings and comments of the experts with their identities removed, the judge model's answers and the analysis scripts.

\begin{credits}
\subsubsection{\ackname} This work was supported by CONVERGENCE of Humans and Machines (220025) and the EVIL-AI ``The identification and the mitigation of the negative effects of Artificial Intelligence Agents'' (JAES/2024/EVIL-AI) projects by Jane and Aatos Erkko Foundation and the ``Multifaceted ripple effects and limitations of human-AI interplay at work, business and society (SYNTHETICA)'' project (358714) by Research Council of Finland. We thank the Conselho Nacional de Desenvolvimento Cient\'ifico e Tecnol\'ogico CNPq (Grant 300883/2025-0; 311533/2025-6).
\end{credits}

\bibliographystyle{splncs04}
\bibliography{referencias}

@misc{act2024eu,
  author       = {{European Parliament and Council of the European Union}},
  title        = {Regulation ({EU}) 2024/1689 of the {European} {Parliament} and of the {Council} of 13 {June} 2024 laying down harmonised rules on artificial intelligence and amending {Regulations} ({EC}) {No} 300/2008, ({EU}) {No} 167/2013, ({EU}) {No} 168/2013, ({EU}) 2018/858, ({EU}) 2018/1139 and ({EU}) 2019/2144 and {Directives} 2014/90/{EU}, ({EU}) 2016/797 and ({EU}) 2020/1828 ({Artificial} {Intelligence} {Act})},
  howpublished = {Official Journal of the European Union, L series, 2024/1689, 12.7.2024},
  year         = {2024},
  note         = {{CELEX} 32024R1689, ELI \url{http://data.europa.eu/eli/reg/2024/1689/oj}}
}

@article{hevner2004design,
  title     = {Design science in information systems
               research},
  author    = {Hevner, Alan R. and
               March, Salvatore T. and
               Park, Jinsoo and
               Ram, Sudha},
  journal   = {MIS Quarterly},
  volume    = {28},
  number    = {1},
  pages     = {75--105},
  year      = {2004},
  doi       = {10.2307/25148625}
}

@book{wohlin2012experimentation,
  author    = {Wohlin, Claes and Runeson, Per and H{\"o}st, Martin and
               Ohlsson, Magnus C. and Regnell, Bj{\"o}rn and Wessl{\'e}n, Anders},
  title     = {Experimentation in Software Engineering},
  publisher = {Springer},
  address   = {Berlin, Heidelberg},
  year      = {2012},
  isbn      = {978-3-642-29043-5},
  doi       = {10.1007/978-3-642-29044-2}
}

@article{galli2026approaching,
  author    = {Galli, Federico and Raulino Dal Pont, Thiago and Sartor, Galileo and
               Contissa, Giuseppe},
  title     = {Approaching the {AI} {Act}... with {AI}: {LLMs} and knowledge graphs to
               extract and analyse obligations},
  journal   = {Computer Law \& Security Review},
  volume    = {60},
  pages     = {106230},
  year      = {2026},
  doi       = {10.1016/j.clsr.2025.106230}
}

@inproceedings{raulino2025lost,
  author    = {Dal Pont, Thiago and Galli, Federico and Sartor, Galileo and
               Contissa, Giuseppe},
  title     = {Lost in {EU} Regulation? {Don't} Worry, {AI} Found the Obligation:
               Extracting and Representing Legal Obligations in the {GDPR} the {DSA}
               and the {AI} {Act}},
  booktitle = {Proceedings of the 20th International Conference on Artificial
               Intelligence and Law (ICAIL 2025)},
  publisher = {ACM},
  year      = {2025},
  pages     = {129-138},
  doi       = {10.1145/3769126.3769260}
}

@article{hayes2007answering,
  author    = {Hayes, Andrew F. and Krippendorff, Klaus},
  title     = {Answering the Call for a Standard Reliability Measure for Coding Data},
  journal   = {Communication Methods and Measures},
  volume    = {1},
  number    = {1},
  pages     = {77-89},
  year      = {2007},
  doi       = {10.1080/19312450709336664}
}

@misc{ralph2026irr,
  author       = {Ralph, Paul and others},
  title        = {{ACM SIGSOFT} Empirical Standards, Inter-Rater Reliability and
                  Agreement supplement},
  howpublished = {ACM SIGSOFT Empirical Standards, GitHub},
  url          = {https://github.com/acmsigsoft/EmpiricalStandards/blob/554118c/docs/supplements/InterRaterReliabilityAndAgreement.md},
  year         = {2026},
  note         = {Commit 554118c of 17 September 2026}
}

@article{krippendorff2013commentary,
  author    = {Krippendorff, Klaus},
  title     = {Commentary: A Dissenting View on So-Called Paradoxes of Reliability
               Coefficients},
  journal   = {Annals of the International Communication Association (Communication
               Yearbook 36)},
  volume    = {36},
  number    = {1},
  pages     = {481-499},
  year      = {2013},
  doi       = {10.1080/23808985.2013.11679143}
}

@inproceedings{abualhaija2025xtrareg,
  author    = {Abualhaija, Sallam and Ceci, Marcello and Sannier, Nicolas and Bianculli, Domenico and Lannier, Salom{\'e} and Siclari, Martina and Voordeckers, Olivier and Tosza, Stanis{\l}aw},
  title     = {{LLM}-assisted Extraction of Regulatory Requirements: A Case Study on the {GDPR}},
  booktitle = {2025 IEEE 33rd International Requirements Engineering Conference (RE)},
  pages     = {142--154},
  year      = {2025},
  doi       = {10.1109/RE63999.2025.00023}
}

@inproceedings{nair2026reg2req,
  author    = {Nair, Pavithra P. M. and Anish, Preethu Rose},
  title     = {From Regulation to Requirements: An Automated Requirement Derivation and Explanation Pipeline},
  booktitle = {2026 IEEE 34th International Requirements Engineering Conference (RE)},
  publisher = {IEEE},
  year      = {2026},
  doi       = {10.1109/RE68928.2026.00038}
}

@article{sovrano2025simplifying,
  author  = {Sovrano, Francesco and Hine, Emmie and Anzolut, Stefano and Bacchelli, Alberto},
  title   = {Simplifying software compliance: {AI} technologies in drafting technical documentation for the {AI} {Act}},
  journal = {Empirical Software Engineering},
  volume  = {30},
  number  = {4},
  pages   = {91},
  year    = {2025},
  doi     = {10.1007/s10664-025-10645-x}
}

@inproceedings{cerqueira2026tere4ai,
  author    = {Siqueira de Cerqueira, Jos{\'e} and Kemell, Kai-Kristian and Waseem, Muhammad and Rousi, Rebekah and Xi, Nannan and Hamari, Juho and Abrahamsson, Pekka},
  title     = {{TERE4AI}: A Knowledge Graph-Based Tool for Generating {EU} {AI} {Act} Compliant Requirements},
  booktitle = {Joint Proceedings of REFSQ-2026 Workshops, Doctoral Symposium, Posters \& Tools Track, and Education and Training Track},
  series    = {CEUR Workshop Proceedings},
  volume    = {4208},
  year      = {2026},
  url       = {https://ceur-ws.org/Vol-4208/pt-short4.pdf}
}

@article{baltes2025guidelines,
  author  = {Baltes, Sebastian and Angermeir, Florian and Arora, Chetan and others},
  title   = {Guidelines for empirical studies in software engineering involving large language models},
  journal = {Empirical Software Engineering},
  volume  = {32},
  pages   = {64},
  year    = {2027},
  doi     = {10.1007/s10664-026-10922-3}
}

@misc{omnibus2026,
  author       = {{European Parliament and Council of the European Union}},
  title        = {Regulation ({EU}) 2026/1744 of the {European} {Parliament} and of the {Council} of 8 {July} 2026 amending {Regulations} ({EU}) 2024/1689, ({EU}) 2018/1139 and ({EU}) 2023/1230 as regards the simplification of the implementation of harmonised rules on artificial intelligence ({Digital} {Omnibus} on {AI})},
  howpublished = {Official Journal of the European Union, L series, 2026/1744, 24.7.2026},
  year         = {2026},
  note         = {{CELEX} 32026R1744; in force 27 July 2026}
}

@misc{hleg2019,
  author = {{High-Level Expert Group on Artificial Intelligence}},
  title  = {Ethics Guidelines for Trustworthy {AI}},
  year   = {2019},
  note   = {{European} Commission, 8 April 2019},
  doi    = {10.2759/346720}
}

@article{runeson2009guidelines,
  author  = {Runeson, Per and H{\"o}st, Martin},
  title   = {Guidelines for conducting and reporting case study research in software engineering},
  journal = {Empirical Software Engineering},
  volume  = {14},
  number  = {2},
  pages   = {131--164},
  year    = {2009},
  doi     = {10.1007/s10664-008-9102-8}
}

@article{zicari2021zinspection,
  author  = {Zicari, Roberto V. and Brodersen, John and Brusseau, James and others},
  title   = {{Z-Inspection}: A Process to Assess Trustworthy {AI}},
  journal = {IEEE Transactions on Technology and Society},
  volume  = {2},
  number  = {2},
  pages   = {83--97},
  year    = {2021},
  doi     = {10.1109/TTS.2021.3066209}
}

@article{kosenkov2025mapping,
  author  = {Kosenkov, Oleksandr and Elahidoost, Parisa and Gorschek, Tony and Fischbach, Jannik and Mendez, Daniel and Unterkalmsteiner, Michael and Fucci, Davide and Mohanani, Rahul},
  title   = {Systematic mapping study on requirements engineering for regulatory compliance of software systems},
  journal = {Information and Software Technology},
  volume  = {178},
  pages   = {107622},
  year    = {2025},
  doi     = {10.1016/j.infsof.2024.107622}
}

@inproceedings{otto2007addressing,
  author    = {Otto, Paul N. and Ant{\'o}n, Annie I.},
  title     = {Addressing Legal Requirements in Requirements Engineering},
  booktitle = {15th IEEE International Requirements Engineering Conference (RE 2007)},
  pages     = {5--14},
  year      = {2007},
  doi       = {10.1109/RE.2007.65}
}

@misc{cerqueira2025mapping,
  author        = {Siqueira de Cerqueira, Jos{\'e} Antonio and Kemell, Kai-Kristian and Rousi, Rebekah and Xi, Nannan and Hamari, Juho and Abrahamsson, Pekka},
  title         = {Mapping Trustworthiness in Large Language Models: A Bibliometric Analysis Bridging Theory to Practice},
  year          = {2025},
  eprint        = {2503.04785},
  archivePrefix = {arXiv},
  doi           = {10.48550/arXiv.2503.04785}
}

@inproceedings{hassani2024rethinking,
  author    = {Hassani, Shabnam and Sabetzadeh, Mehrdad and Amyot, Daniel},
  title     = {Rethinking Legal Compliance Automation: Opportunities with Large Language Models},
  booktitle = {2024 IEEE 32nd International Requirements Engineering Conference (RE)},
  pages     = {432--440},
  publisher = {IEEE},
  year      = {2024},
  doi       = {10.1109/RE59067.2024.00051}
}

@inproceedings{panickssery2024llm,
  author    = {Panickssery, Arjun and Bowman, Samuel R. and Feng, Shi},
  title     = {{LLM} Evaluators Recognize and Favor Their Own Generations},
  booktitle = {Advances in Neural Information Processing Systems 37 (NeurIPS 2024)},
  year      = {2024},
  doi       = {10.52202/079017-2197}
}

@inproceedings{cerqueira2025trust,
  author    = {Siqueira de Cerqueira, Jos{\'e} Antonio and Agbese, Mamia and Rousi, Rebekah and Xi, Nannan and Hamari, Juho and Abrahamsson, Pekka},
  title     = {Can We Trust {AI} Agents? A Case Study of an {LLM}-Based Multi-Agent System for Ethical {AI}},
  booktitle = {Proceedings of the Conference on Technology Ethics 2025 (Tethics 2025)},
  series    = {CEUR Workshop Proceedings},
  volume    = {4237},
  pages     = {68--82},
  year      = {2025},
  url       = {https://ceur-ws.org/Vol-4237/paper6.pdf}
}

@inproceedings{sillberg2025falls,
  author    = {Veesommai Sillberg, Chalisa and Siqueira De Cerqueira, Jos{\'e} and Sillberg, Pekka and Kemell, Kai-Kristian and Abrahamsson, Pekka},
  title     = {The {EU} {AI} {Act} is a Good Start But Falls Short},
  booktitle = {Software Business (ICSOB 2024)},
  series    = {LNBIP},
  volume    = {539},
  pages     = {114--130},
  publisher = {Springer},
  year      = {2025},
  doi       = {10.1007/978-3-031-85849-9_10}
}

\end{document}